# Buckling activated translation of encapsulated microbubbles

Maria Vlachomitrou[1], Georges Chabouh[2], Alkmini Lytra[1] and Nikos Pelekasis[1*]

[1] Department of Mechanical Engineering, University of Thessaly, Leoforos Athinon, Pedion Areos, 38334 Volos, Greece

[2] Department of Biomedical Engineering, Department of Radiology, Columbia University, New York, NY, USA

**Abstract**

The influence of initial shape imperfections on the post-buckling and translational behavior of encapsulated microbubbles is investigated subject to acoustic excitation in an unbounded flow. Bifurcation analysis reveals that imperfections lower the critical overpressure for buckling, guiding the bubble along stable branches with transitions influenced by the nature of the imperfections. The resulting branches emerge as unfoldings of bifurcation diagrams and closely resemble solution families emanating from the standard bifurcation diagram of a perfect sphere. Dynamic simulations for fixed forcing frequency demonstrate that the microbubble buckles at the defective pole and at maximum compression adopts the shape corresponding to the buckled shape that emerges in the static bifurcation diagram for the same pressure disturbance. Simulations further reveal that the deformed microbubble exhibits translational motion in the direction of concavity, driven by the balance between pressure and viscous drag, both aligned with this motion as a result of the deformed shape acquired by the microbubble. The translational speed is highly dependent on the shell stretching and bending elasticities and forcing frequency, with an optimal speed achieved when the forcing frequency equals the difference between the resonance frequencies of the initial shape and the compressed static shape corresponding to the imposed sound amplitude. Higher sound amplitude or lower bending resistance lead to more pronounced deformations that enhance translational motion. The impact of this dynamic response pattern on the recently reported translational motion of properly engineered coated microbubbles to act as microswimmers in modern drug delivery modalities under ultrasound excitation, is highlighted.

* Correspondence: Nikos Pelekasis, pel@uth.gr; Tel.: (+30)24210-74102

## I. INTRODUCTION

Biological microswimmers, such as bacteria, are microorganisms capable of moving autonomously in liquid environments. In these environments, viscous damping typically dominates over inertia. However, microorganisms have evolved propulsion strategies that overcome or even exploit viscous drag [1]. This remarkable ability of microorganisms has inspired scientists to explore ways of building artificial microswimmers for use in various technical and medical applications. The potential applications of microswimmers in field such as drug delivery, assisted fertilization, imaging and other biomedical areas are highly promising and are currently being actively explored [2]. Different types of artificial microswimmers have been proposed, some of which rely on external sources for propulsion. Examples include magnetic-driven [3] and acoustic-driven swimmers [4, 5]. Others, such as catalytic microswimmers, move autonomously [6]. Additionally, research is exploring the combination of various mechanisms to enhance functionality.

While the use of microswimmers in biomedical applications holds considerable promise, several challenges must still be addressed before they can be effectively implemented. These include ensuring the biocompatibility and biodegradability of the materials, as well as developing efficient navigation mechanisms within the complex human fluid environment. In this context, ultrasound emerges as a promising driving mechanism, owing to its widespread use in medical applications. Additionally, interaction between microbubbles and sound waves has attracted interest due to their potential as auxiliary propulsion systems using acoustic radiation forces for drug delivery purposes.

In a pioneering study, Dijkink et al. [7] developed a device known as the "acoustic scallop," which moves due to a quasi-steady streaming flow. The device consists of a small tube that is sealed at one end and contains a gas bubble submerged in a liquid. When a sound field is applied, the bubble oscillates causing the fluid to move in and out of the open end of the tube. Provided that the inertia of the flow exiting the tube is significant and the Reynolds number is not too low [8], an asymmetrical flow pattern is generated, resulting in a jet of fluid being expelled outward. This creates a net momentum over a cycle, propelling the device forward. In their experiments Dijkink et al. [7] achieved a maximum velocity of 1.35 mm/s for a device with a 750 μm diameter and 2–4 mm length.

Building on this concept, Feng et al. [9] tried to reduce the device size to micro-scale using microphotolithography, fabricated 2D-microswimmers and reported a maximum velocity of 40 mm/s. To counteract the reduced Reynolds number, and consequently the weaker propulsion force as the device size decreases, they tuned the sound frequency near the resonance frequency of the swimmer. Other lines of research focused on 3D microswimmers. For instance, Bertin et al. [4] introduced armored microbubbles (AMBs) in place of gas bubbles in order to delay gas dissolution, thus achieving translational velocities up to 100 mm/s. Meanwhile, Louf et al. [10] designed 3D microswimmers with trapped air oriented towards the substrate and reported a maximum velocity of 350 mm/s. Their work highlighted the significant role of frequency and sound amplitude in swimmer motion, suggesting that two acoustic forces are crucial for navigation: the acoustic streaming force, which helps the swimmer overcome adhesion forces and hover, and the radiation force which drives motion towards the substrate.

Building on the fact that encapsulated microbubbles, commonly known as ultrasound contrast agents, are already approved for use in biomedical applications, Chabouh et al. [11] explored the possibility of utilizing them as swimming agents instead of conventional gas bubbles. Contrast agents are widely used in ultrasound imaging of vital organs, as their non-linear response to sound enhances ultrasound backscatter and contrast, relative to surrounding tissue, while producing high-quality images [12] and super-resolution ultrasound images [13]. Beyond imaging, as was above mentioned, contrast agents are also considered for targeted drug delivery [14]. In this case, encapsulated microbubbles carrying drugs are directed and attached to the affected site using appropriate acoustic disturbances. The drug is then released locally either by rupturing the shell via sonication [15, 16] or by subjecting the microbubble to low amplitude oscillations that facilitate drug delivery by generating transient micropores in nearby cells via cyclic microjets [17, 18] or via the acoustic microstreaming process [19].

In their study, Chabouh et al. [11] subjected commercial contrast agents SonoVue® (Bracco, Italy) and homemade contrast agents [20] to external sinusoidal pressure cycles and reported a net displacement via repeated, non-destructive cycles of compression and expansion of the microbubble. In particular, the microbubbles exhibited oscillations between an almost spherical shape during expansion and an asymmetric buckled bowl-like shape during compression. Notably, motion was observed in the direction of the pole where buckling occurred. The location around which buckling develops (buckling spot) and, thus, the direction of the motion were linked to the presence of a defect on the shell that was noticeable, but its size could not be accurately determined. In a related study, Djellouli et al. [21] successfully investigated buckling instability as a swimming mechanism to propel millimetric colloidal shells filled with air. Their findings showed that regardless of the shell's thickness or the fluid viscosity, the shell consistently moved in the direction dictated by the buckling pattern.

The influence of shape asymmetry on the direction of motion of an oscillating contrast agent was examined in a different context in the work of Vlachomitrou & Pelekasis [22]. The latter study focused on the translational motion of contrast agents in the presence of a nearby boundary, specifically after the onset of the compression-only behavior, i.e. oscillations with the compression phase being much more intense than the expansion phase. It was demonstrated that the concave direction of the compressed buckled shape around which the bubble oscillates aligns with the direction of bubble movement. Numerical simulations revealed that when concave upward compressed shapes occur, the bubble's entrapment near the underlying boundary is interrupted and a motion in the opposite direction is initiated due to the positive pressure drag that develops. Conversely, symmetric shapes or asymmetric shapes that are concave downward continued translating toward the boundary under the influence of secondary Bjerknes forces.

The similarities between the experiments of Chabouh et al. [11] and the simulations of Vlachomitrou & Pelekasis [22], notably the reported asymmetric buckled shapes and the observation that the buckled shape seems to determine the direction of motion, motivated the current study. In this work, we aim to numerically investigate the mechanism responsible for the motion of an encapsulated bubble with a small shell defect in an unbounded flow. Additionally, we seek to define, where possible, the conditions that maximize the translational velocity of the bubble.

This paper is organized as follows: The problem formulation is presented in Sec. II, while the numerical methodology is outlined in Sec. III and results are presented and discussed in Sec. IV. Section IV.A provides the bifurcations diagrams, highlighting the effect of the shell defect, and Sec. IV.B focuses on numerical simulations and the identification of the driving mechanism of the "swimming" motion. Finally, in Sec. V the main conclusions are summarized and practical implications are discussed.

## II. PROBLEM FORMULATION

We are interested in examining the dynamic response to acoustic disturbances of an encapsulated microbubble that bears a small defect on the shell. The bubble is initially stress-free, with an equivalent radius, $R_{0,eq}$, pertaining to a spherical bubble that occupies the same volume, $V_0$, as the initial bubble with the defect and is submerged in a Newtonian liquid of density $\rho$ and dynamic viscosity $\mu$. We consider an unbounded flow, and we investigate the bubble's response to an acoustic signal imposed on the far pressure field:

$$P_\infty = P_{st} + P_{dist} = P_{st} + P_{st}\varepsilon\cos(t), \tag{1}$$

with $P_{st}, P_{dist}$ denoting the dimensionless undisturbed and disturbed pressure in the far field, respectively, and $\varepsilon$ the amplitude of the acoustic disturbance. The undisturbed pressure in the far field is fixed to the standard atmospheric pressure, 101.325 kPa.

To describe the problem, we employ the spherical coordinate system, and we assume axisymmetric variations of the bubble shape as well as the liquid velocity and pressure. In Fig. 1 a schematic representation of the flow under consideration is provided, with the small defect placed at the pole $\theta=0$ and $f_1$ denoting the r-coordinate of the shell that coats the bubble.

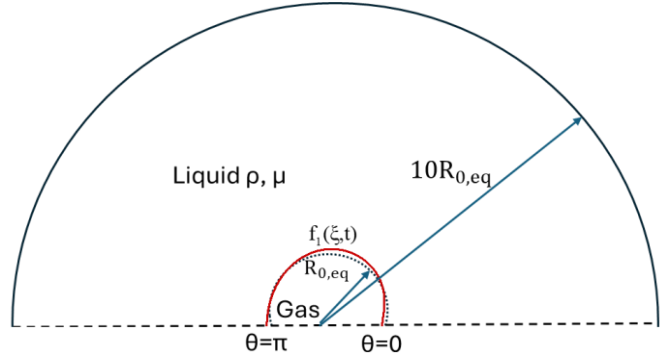

FIG. 1. An encapsulated microbubble with a shape defect at θ=0 in an unbounded flow.

The initial equivalent radius, $R_{0,eq}$, of the bubble is considered as the characteristic length scale, whereas the appropriate time scale is set to $1/\omega_f$, with $\omega_f$ signifying the angular forcing frequency of the pressure disturbance. In this context, the characteristic velocity and pressure scales are equal to $\omega_f R_{0,eq}$ and $\rho\omega_f^2 R_{0,eq}^2$, respectively.

The flow in the surrounding incompressible liquid is governed by the continuity and Navier-Stokes equations that read in dimensionless form:

$$\nabla \cdot \boldsymbol{u} = 0, \tag{2}$$

$$\frac{\partial \boldsymbol{u}}{\partial t} + (\boldsymbol{u}\cdot\nabla)\boldsymbol{u} = -\nabla P + \frac{1}{Re}\nabla\cdot\underline{\underline{\boldsymbol{\tau}_l}}, \quad \underline{\underline{\boldsymbol{\sigma}}} = -P\underline{\underline{\boldsymbol{I}}} + \frac{1}{Re}\underline{\underline{\boldsymbol{\tau}_l}}, \quad \underline{\underline{\boldsymbol{\tau}_l}} = \nabla\boldsymbol{u} + \nabla\boldsymbol{u}^T, \tag{3}$$

where $\boldsymbol{u}=(u_r,u_\theta,0)$, $Re=(\rho\omega_f R_{0,eq}^2)/\mu$ is the Reynolds number of the flow that compares inertia with viscous forces, $\underline{\underline{\boldsymbol{\sigma}}}, \underline{\underline{\boldsymbol{\tau}_l}}$, the full and deviatoric stress tensors in the surrounding fluid and $\underline{\underline{\boldsymbol{I}}}$ the unit tensor.

A Lagrangian representation is employed to describe the shape of the bubble by introducing the coordinate $\xi$ ($0\leq\xi\leq 1$), with $\xi=0$ and $\xi=1$ corresponding to the poles $\theta=0$ and $\theta=\pi$, respectively. The force balance on the gas-liquid interface is given by:

$$\left(-P\underline{\underline{\boldsymbol{I}}}+\frac{1}{Re}\underline{\underline{\boldsymbol{\tau}_l}}\right)\cdot\boldsymbol{n}+P_G\boldsymbol{n} = -\nabla_s\cdot(\underline{\underline{\boldsymbol{\tau}}}+q\boldsymbol{n})+\frac{2k_m}{We}\boldsymbol{n} = \boldsymbol{\Delta F}+\frac{2k_m}{We}\boldsymbol{n} \tag{4}$$

where $\mathbf{n} = (r\theta_\xi / \sqrt{r_\xi^2 + r^2\theta_\xi^2})\mathbf{e}_r - (r_\xi / \sqrt{r_\xi^2 + r^2\theta_\xi^2})\mathbf{e}_\theta$ is the unit normal vector pointing towards the surrounding fluid, $P_G$ the gas pressure inside the bubble, $\nabla_s, k_m$ the surface gradient and mean curvature of the bubble interface, respectively, $\mathbf{q}$ the transverse shear force resultant and $We = \dfrac{\rho \omega_f^2 R_{0,eq}^3}{\sigma}$ the Weber number comparing inertia with capillary forces. Following our previous studies, we set $\sigma = 0.051$ N/m as a measure of the internal gas exposure to the surrounding liquid. Finally, the term $\mathbf{\Delta F}$ in Eq. (4) refers to the resultant force due to the viscoelastic properties of the membrane and is equal to:

$$\mathbf{\Delta F} = \left[ k_s \tau_s + k_\phi \tau_\phi - \frac{1}{r_o} \frac{\partial}{\partial s}(r_o q) \right] \mathbf{n} - \left[ \frac{\partial \tau_s}{\partial s} + \frac{1}{r_o} \frac{\partial r_o}{\partial s}(\tau_s - \tau_\phi) + k_s q \right] \mathbf{e}_s, \qquad (5a)$$

$$q = \frac{K_B}{r_o} \frac{\partial r_o}{\partial s} \left[ \frac{\partial}{\partial r_o}(r_o m_s) - m_\phi \right], \qquad K_B = \frac{k_B}{\rho \omega_f^2 R_{0,eq}^5}, \qquad (5b)$$

with $s, \phi$ denoting the interfacial arclength and azimuthal directions, respectively, $\tau_s, \tau_\phi$ the principal stress resultants, $k_s, k_\varphi$ the two principal curvatures, $m_s, m_\varphi$ the principal bending moments, $k_B$ the bending modulus, $r_o = r\sin\theta$ the cylindrical polar coordinate and $\mathbf{e}_s$ the tangential unit vector. In Eqs. (5) $q$ refers to the transverse shear tension that is obtained from a torque balance on the shell [23, 24] and $K_B$ signifies the relative importance of bending with respect to inertia. The membrane and bending stresses are defined via the shell constitutive laws. The principal membrane tensions consist of an elastic, $\tau_{el}$, and a viscous component, $\tau_v$, e.g. $\tau_s = \tau_{s,el} + \tau_{s,v}$. For the elastic part, we adopt the Mooney–Rivlin constitutive law [26] with the degree of softness $b$ set to zero [25, 26]:

$$\tau_{s,el} = \frac{G}{3\lambda_s \lambda_\varphi}\left( \lambda_s^2 - \frac{1}{(\lambda_s \lambda_\varphi)^2} \right)\left[ 1 + b\left(\lambda_\varphi^2 - 1\right) \right], \qquad (6a)$$

$$\tau_{\phi,el} = \frac{G}{3\lambda_s \lambda_\varphi}\left(\lambda_\phi^2 - \frac{1}{(\lambda_s \lambda_\varphi)^2}\right)\left[1+b(\lambda_s^2-1)\right], \tag{6b}$$

where $G = \chi/(\rho\omega_f^2 R_{0,eq}^3)$ compares the importance of shell dilatation with respect to inertia and $\lambda_s$, $\lambda_\varphi$ correspond to the principal extension ratios based on the stress-free state:

$$\lambda_s = \frac{s_\xi(t)}{s_\xi(0)}, \qquad \lambda_\phi = \frac{r(t)\sin\theta(t)}{r(0)\sin\theta(0)}, \qquad s_\xi = \sqrt{r_\xi^2 + r^2\theta_\xi^2}. \tag{7}$$

For the viscous part, the dilatational viscosity $\mu_s$ and shear viscosity $\mu_{sh}$ of the shell are treated separately and, thus, the viscous components of the membrane tensions are given by:

$$\tau_s^v = \left(\frac{1}{Re_s}+\frac{1}{Re_{sh}}\right)\frac{1}{\lambda_s}\frac{\partial\lambda_s}{\partial t} + \left(\frac{1}{Re_s}-\frac{1}{Re_{sh}}\right)\frac{1}{\lambda_\varphi}\frac{\partial\lambda_\varphi}{\partial t}, \tag{8a}$$

$$\tau_\phi^v = \left(\frac{1}{Re_s}+\frac{1}{Re_{sh}}\right)\frac{1}{\lambda_\phi}\frac{\partial\lambda_\phi}{\partial t} + \left(\frac{1}{Re_s}-\frac{1}{Re_{sh}}\right)\frac{1}{\lambda_s}\frac{\partial\lambda_s}{\partial t}, \tag{8b}$$

with $Re_s = \rho\omega_f R_{0,eq}^3/\mu_s$ and $Re_{sh} = \rho\omega_f R_{0,eq}^3/\mu_{sh}$ comparing inertia forces with the viscous dilatational and shear forces of the shell, respectively. A more detailed description of the modelling of the viscoelastic shell of the bubble can be found in Tsiglifis & Pelekasis [26-28] and in Pelekasis et al. [29].

The continuity of the liquid and shell velocities is also applied on the interface:

$$\mathbf{u} = \frac{D\mathbf{r}_s}{Dt} \Rightarrow \quad u_r = \frac{\partial r}{\partial t}, \quad u_\theta = r\frac{\partial\theta}{\partial t}, \tag{9}$$

with $\mathbf{r}_s = r\mathbf{e}_r$ corresponding to the position vector of a particle at the interface.

The gas pressure inside the bubble is considered uniform due to negligible density and kinematic viscosity. Moreover, heat transfer between the bubble and the surrounding liquid is assumed to take place fast in comparison with the time scale of the phenomena under consideration. In this context, bubble oscillations are treated as nearly isothermal, and therefore the bubble pressure is given by:

$$P_G(t=0)V_G^\gamma(t=0) = P_G(t)V_G^\gamma(t) = const., \tag{10}$$

with $V_G$ being the dimensionless instantaneous volume of the bubble, $V_G(t=0)$ the initial volume of the bubble and $\gamma$ the polytropic constant set to 1.07.

Initially a stress-free state is assumed with a shape characterized by a geometric imperfection in the radial distance from the center of mass in the form of an axisymmetric radial depression distribution $w(\theta)$ around the pole, of the form:

$$w(\theta) = w_0 \exp\left(-\frac{\theta}{\theta_0}\right) \tag{11}$$

where $w_0$ indicates the maximum depression and $\theta_0$ the range of azimuthal angles $\theta$ along the interface affected by the imperfection; $r$, $\theta$ signify the radial and azimuthal spherical coordinates, respectively. In this context, considering a spherical microbubble with radius $R_{sph}$ and performing a mass balance between the reference spherical state and the stress-free state with the imperfection

$$P_G(t=0)V_G^\gamma(t=0) = P_{G,Sph}V_{G,Sph}^\gamma, \tag{12}$$

we obtain an estimate for the internal pressure at static equilibrium for the microbubble with the imperfection and a standard atmospheric external pressure, $P_{st}$; e.g. for a reference radius of $R_{sph}$= 4 μm we obtain an imperfect bubble with an equivalent radius of $R_{0,eq}\approx$3.996 μm. The approach with an initial depression of the form shown in Eq. (11) was also applied to the static arrangement with an imperfection on each one of the two poles, Figs. 2(a), 2(b), in which case the equivalent initial radius was slightly lower than the above value, with the corresponding larger initial pressure. The latter conditions constitute the initial stress-free state of the microbubble. Upon this base state, either a uniform pressure change is applied in order to construct the bifurcation diagrams arising at static equilibrium, or an acoustic signal is imposed on the far field in order to investigate the dynamic response of the microbubble.

## III. NUMERICAL METHODOLOGY

The static equilibrium for a stress-free state characterized by a slight indentation, as illustrated in Fig. 2, is solved numerically in the manner described in Lytra & Pelekasis [30] and Pelekasis et al. [29] and the emerging bifurcation diagrams were constructed so as to facilitate the interpretation of the ensuing dynamic simulations and the identification of the mechanism responsible for the onset of the translational motion. In particular, the force balance and isothermal pressure variation equations are discretized with the finite element method using the B-cubic splines as basis functions and the non-linear system is solved iteratively with the Newton-Raphson method. The stability of each solution branch is studied by calculating the corresponding eigenvalues using the dgeev Lapack subroutine. Simple or arc-length continuation is performed to follow the evolution of the solution in monotonic and limit point areas, respectively, as the external pressure varies. When the bifurcation diagram of a perfect sphere is considered, the post-buckling curves are constructed by disturbing the compressed spherical shape at the bifurcation point with the corresponding eigenvector. In contrast, the transition to the buckled shapes is spontaneous without disturbing any solution in the case of an imperfection on the initial stress-free shape.

The numerical methodology employed for the dynamic simulations of the acoustically excited coated microbubble is described in detail in Vlachomitrou and Pelekasis [31] and Pelekasis et al. [29]. Specifically, we use a superparametric Galerkin finite element formulation based on the standard biquadratic/bilinear representation for the velocity and pressure fields. The fourth-order derivative that arises in the force balance equation, Eqs. (4), (5), due to the bending resistance of the shell necessitates the introduction of B-cubic splines to represent the unknown location of the interface. A fully implicit scheme is used for time integration, while the nonlinearity of the problem is handled via the Newton-Raphson method. The final set of linearized equations is solved iteratively using the GMRES method [32].

The mesh is constructed elliptically, as this method has proven extremely successful in adjusting the grid to accommodate high interfacial distortions, as demonstrated in our previous studies involving both unbounded flows [30, 32] and wall-restricted flows [22, 33, 34]. The grid construction method for the unbounded flow is given in detail in [29, 31] and it is based on solving the following coupled equations, along with appropriate boundary conditions [35, 36]:

$$\nabla \cdot \left( \varepsilon_1 \sqrt{\frac{r_\xi^2 + r^2 \theta_\xi^2}{r_\eta^2 + r^2 \theta_\eta^2}} + 1 - \varepsilon_1 \right) \nabla \xi = 0 \tag{13}$$

$$\nabla \cdot \nabla \eta = 0 \tag{14}$$

These equations are discretized using the biquadratic Lagrangian basis functions. The first equation produces the $\eta$-curves which must intersect the interface almost orthogonally, while the second equation generates the $\xi$-curves which are nearly parallel to the interface. The term $\varepsilon_1$ controls the mesh smoothness vs. orthogonality, ranges between 0 and 1 and is set to 0.1 throughout this paper.

As an overall numerical procedure, at each time step two separate Newton-Raphson procedures are applied. First, the governing equations presented in Sec. II are solved simultaneously to obtain the velocity and pressure fields along with the shape of the interface. The resultant interfacial shape is then imposed as an essential boundary condition for the second Newton-Raphson iterative procedure, which constructs the updated grid. Details regarding the discretized form of the governing equations, the boundary conditions, and the implementation of the Newton-Raphson, the GMRES and the elliptic mesh generation methods can be found in [31].

## IV. RESULTS AND DISCUSSION

This study is motivated by the work of Chabouh et al. [11], and therefore, we use shell parameters within the range of those employed in their experiments. We examined two cases of phospholipid shell-coated microbubbles.

- **Case I** refers to an initially stress-free bubble with an equivalent radius of $R_{0,eq}$ = 3.996μm, shell viscosity $\mu_s = \mu_{sh} = 3.144 \cdot 10^{-9}$ Kg/s, area dilatation modulus $\chi$=2 N/m, and bending modulus $k_b = 7.4 \cdot 10^{-14} N \cdot m$.

- **Case II** (SonoVue) refers to an initially stress-free bubble with an equivalent radius of $R_{0,eq}$ = 3.996μm, shell viscosity $\mu_s = \mu_{sh} = 3.144 \cdot 10^{-9}$ Kg/s, area dilatation modulus $\chi$=0.5 N/m and bending modulus $k_b = 1.852 \cdot 10^{-14} N \cdot m$.

The surface tension $\sigma$ is set to 0.051 N/m, the polytropic ideal gas constant to $\gamma$=1.07, the Poisson ratio is taken equal to 0.8 and $k_b / (\chi R_0^2) = (d_{eff} / R_0)^2 / 6 / (1+\nu)$, as proposed in Chabouh et al. [37]. Both microbubbles exhibit a similar small defect located around $\theta$=0, shown in Fig.2(a) with the interfacial shape expressed in terms of the Cartesian coordinates $z = r\cos\theta$ and $x = r\sin\theta$ subject to an initial disturbance of the form given in Eqs. (11), (12).

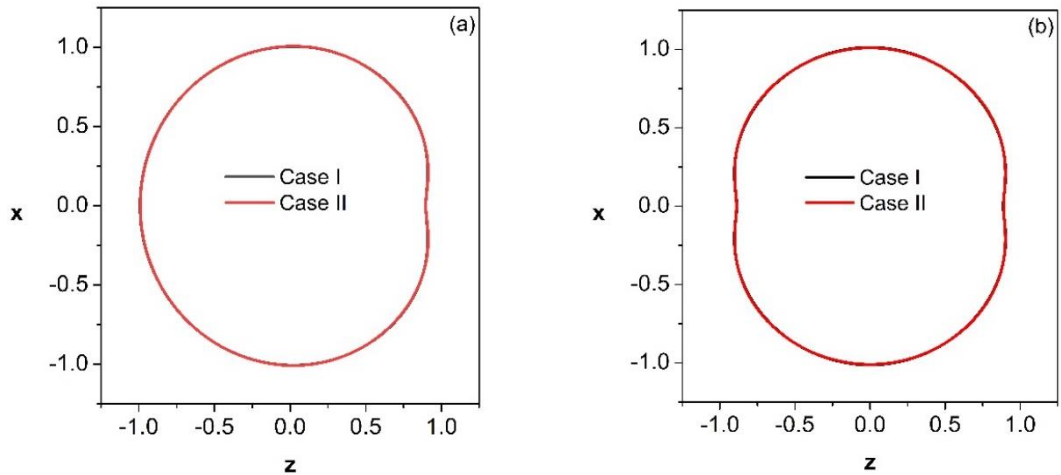

FIG. 2. The initial stress-free bubble for Case I and Case II. (a) Asymmetric and (b) Symmetric imperfection with $w_0$=0.1

## A. Bifurcation diagrams

Initially, we consider the static equilibrium when an increasing uniform overpressure $\varepsilon$ is applied across the coated microbubble.

$$P_\infty = P_{st} + P_{dist} = P_{st} + P_{st}\varepsilon \tag{15}$$

The stress-free reference state may be a perfect sphere or a nearly spherical shell with a small indentation on one or both the poles of the microbubble, see Fig. 2. The latter case is studied to demonstrate the effect of an initial shape imperfection on the bifurcation diagram pertaining to static buckling, compared to that of a perfect spherical shell, with the single imperfection considered as more relevant to the interpretation of experiments with fabricated microswimmers.

When the stress-free shape is a perfect sphere, we obtain gradually compressed spherical solutions that correspond to the black curves in Fig. 3. As the external overpressure $\varepsilon$ increases, buckling modes emerge from the main spherical branch. In both cases, the first bifurcation point is dominated by the $P_4$ Legendre mode ($\varepsilon_{P4} \approx 1.56$ for $\chi=2$ N/m and $\varepsilon_{P4} \approx 0.90$ for $\chi=0.5$ N/m) and the second one by $P_5$ ($\varepsilon_{P5} \approx 1.66$ for $\chi=2$ N/m and $\varepsilon_{P5} \approx 0.95$ for $\chi=0.5$ N/m). The two new solution branches arise (cyan and blue curves in Fig. 3) by disturbing the spherical shape with the corresponding eigenvector at each bifurcation point. The symmetric solution family dominated by $P_4$ is subcritical while the one characterized by the asymmetric $P_5$ Legendre mode is a transcritical branch. Both solution families exhibit a limit point as they evolve towards lower overpressures which turns the branch to higher overpressures as the microbubble is further compressed along the post-buckling curve. The corresponding shapes of each solution branch are depicted in Figs. 4(a)-4(c).

Repeating the same calculations but taking into account an initial indentation on one pole of the bubble for the stress-free shape, i.e. an asymmetric imperfection, we capture a spontaneous transition from the main branch to the buckling branch, see also the green curve in Fig. 3. Interestingly, a symmetric imperfection, as in Fig 2(b), leads to a similar response, where the solution branch now evolves along the $P_4$ post buckling path, red curve in Fig. 3. This behavior constitutes an "unfolding" of the bifurcation diagram corresponding to perfect buckling [39], where the depression distribution in Eq. (11) provides the initial shape of the imperfection around. Moreover, the imperfection is responsible for a significant reduction of the critical pressure that leads to buckling. In fact, the transition from the spherical branch to the buckled curves takes place at around $\varepsilon=0.8$ and $\varepsilon=0.65$ for Case I and II, respectively. During this spontaneous transition the imperfect buckling does not follow the subcritical branch of the post-buckling solution family before the limit point, which is typically characterized by higher total energy [29], but rather it evolves along the most stable path. The latter is more likely to be seen in nature and it is confirmed by the numerical simulations below, as well as by the experimental findings of Chabouh et al. [11, 37]. Shapes pertaining to case II are not shown in Fig. 4 as they do not exhibit a substantially different behavior in terms of deformation and connectivity.

The corresponding shapes for asymmetric and symmetric initial imperfection are depicted in Figs. 5(a), 5(b) and Figs. 6(a), 6(b) for $\chi=2$ N/m and 0.5, respectively, where the exhibited indentation is gradually amplified as the overpressure $\varepsilon$ increases. In both cases, the shapes of the softer microbubble are more compressed for the same applied pressure. In addition, as can be gleaned upon inspection of Figs. 4-6, the shapes obtained from the imperfect bifurcation are almost the same as the ones obtained for the perfect sphere after the limit points of the original bifurcating solution families. In the latter parameter range the solution branches are nearly identical, as expected for bifurcation diagrams obtained by imposing a small imperfection on the perfect bifurcation diagram [38].

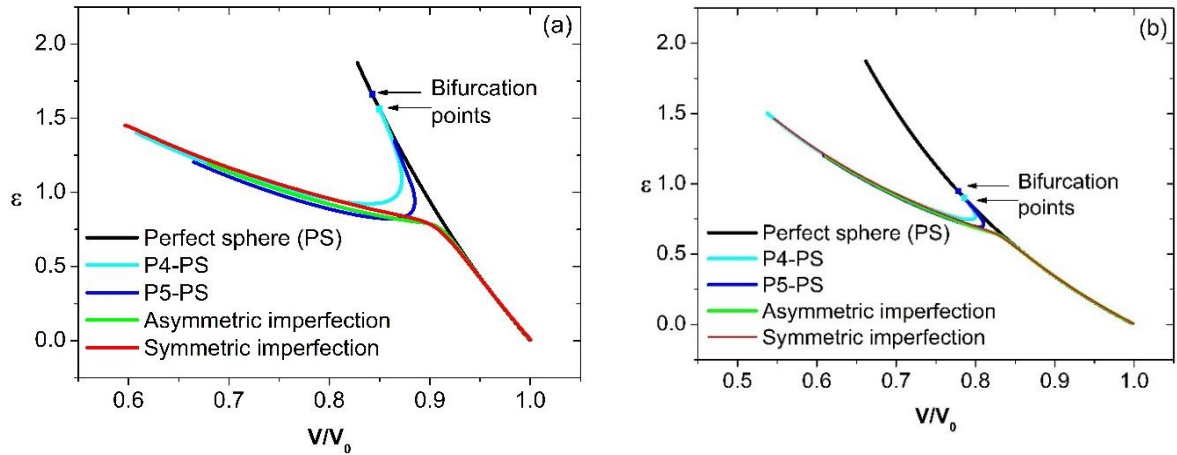

FIG. 3. Bifurcation diagrams of overpressure $\varepsilon$ vs dimensionless volume. (a) Case I and (b) Case II.

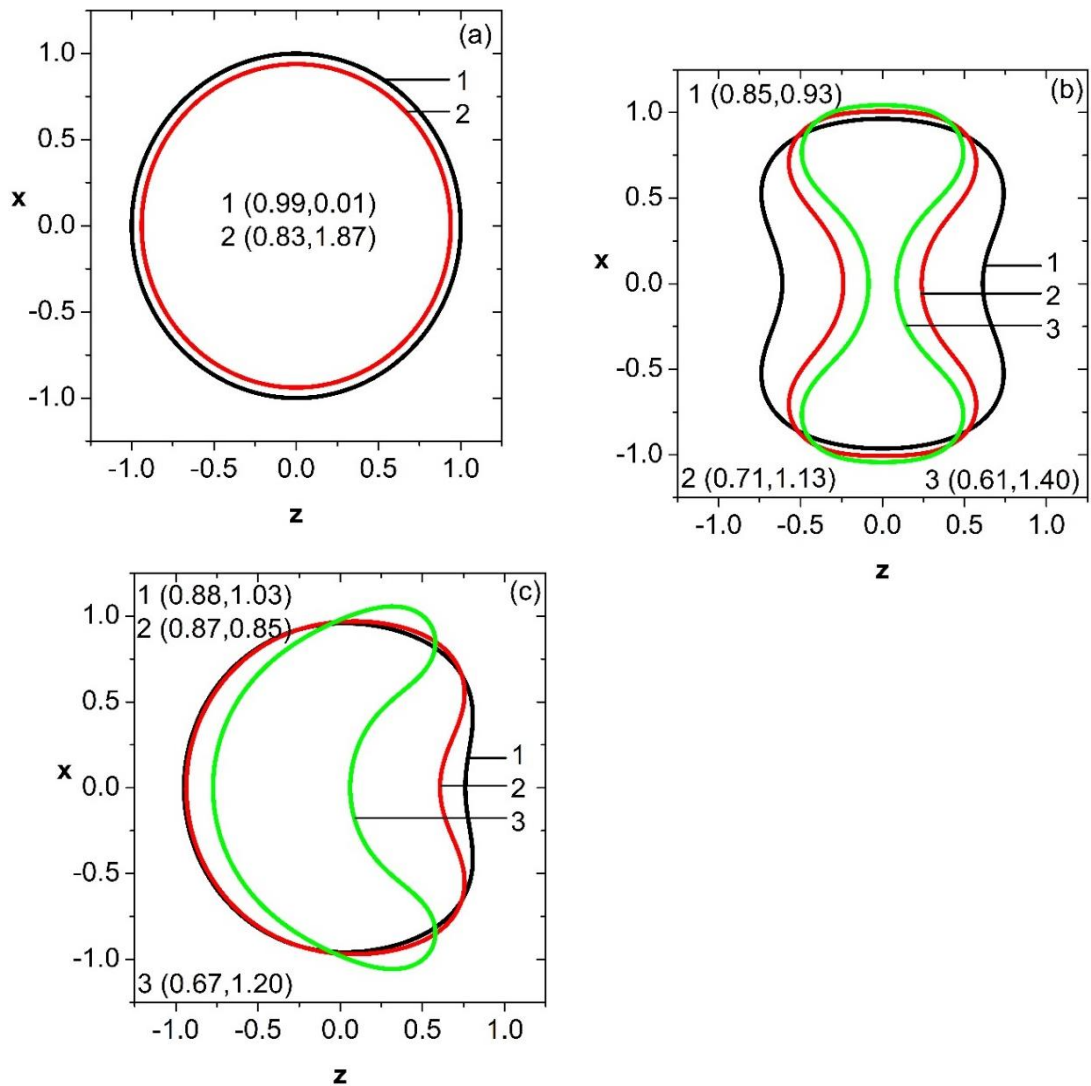

FIG. 4. (a)-(c) Shape of the microbubble for Case I corresponding to $P_0$ spherical, $P_4$ oblate and $P_5$ branches, respectively, for selected points on the bifurcation diagram ($V/V_0$, $\varepsilon$) when the stress-free state is a perfect sphere.

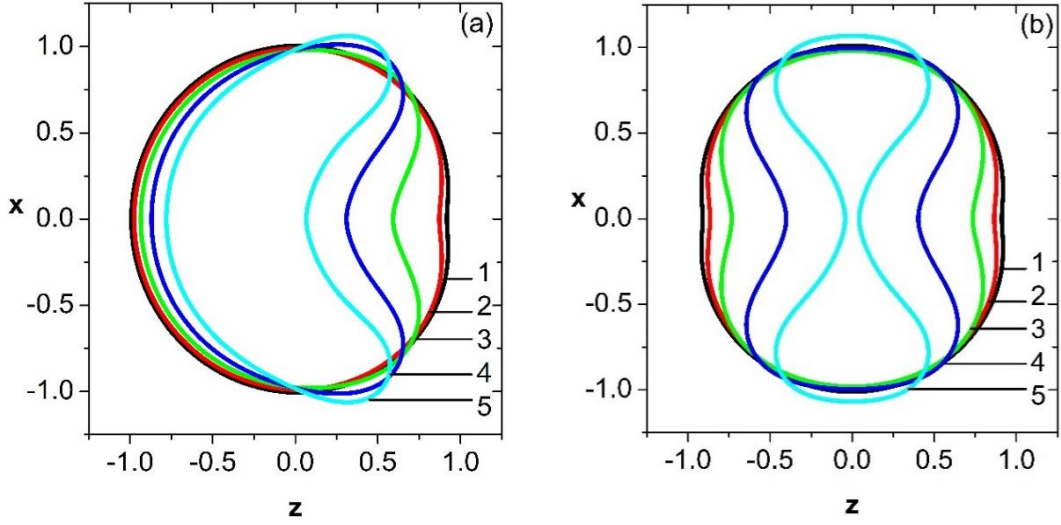

FIG. 5. Shape of the microbubble for Case I with (a) an *asymmetric imperfection* for selected points on the bifurcation diagram ($V/V_0$, $\varepsilon$) = 1. (0.99,0.01), 2. (0.95,0.41), 3. (0.89,0.80), 4. (0.81,0.91), 5. (0.68,1.20) and (b) a *symmetric imperfection* for selected points on the bifurcation diagram ($V/V_0$, $\varepsilon$) = 1. (0.99,0.01), 2. (0.95,0.42), 3. (0.90,0.78), 4. (0.80,0.95), 5. (0.59,1.45).

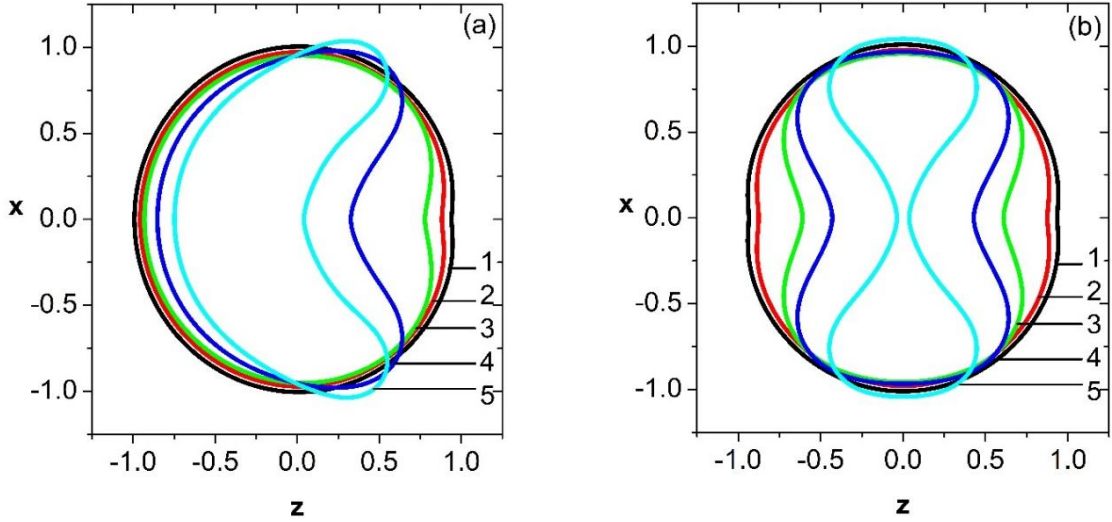

FIG. 6. Shape of the microbubble for Case II with (a) an *asymmetric imperfection* for selected points on the bifurcation diagram ($V/V_0$, $\varepsilon$) = 1. (0.99,0.01), 2. (0.90,0.34), 3. (0.84,0.60), 4. (0.75,0.80), 5. (0.61,1.20) and (b) a *symmetric imperfection* for selected points on the bifurcation diagram ($V/V_0$, $\varepsilon$) = 1. (0.99,0.01), 2. (0.90,0.34), 3. (0.80,0.69), 4. (0.75,0.80), 5. (0.55,1.46).

## B. Numerical simulations

Besides bifurcation diagrams, another useful piece of information that will assist in the interpretation of the dynamic simulations, as well as the available measurements, pertains to the resonance frequency of the acoustically excited microbubble, shown in Fig. 2(a). Since the shape of the bubble when the buckling bifurcation occurs is quite distorted and the Poisson ratio is set to 0.8 instead of the typical value of 0.5, the eigenfrequency cannot be calculated using the formula presented in Vlachomitrou & Pelekasis [33] and Pelekasis et al. [29]. Instead, in the interest of generality, it is obtained dynamically by applying a small step change in pressure, $P_\infty = P_{st}(1+\varepsilon)$, on the initial configuration and determining the frequency of the decaying volume pulsations that the bubble undergoes until it settles to the equilibrium state corresponding to the imposed disturbance. To this end, we fist applied an infinitesimal pressure jump ($\varepsilon = 0.1$) on the stress free static shape with the imperfection around the north pole, and recorded the period of the decaying oscillations pertaining to Cases I and II until static configuration was reached. By performing FFT (Fast Fourier Transform) analysis, illustrated in Fig. 7(a), on the oscillations of the breathing mode, $P_0$, until static equilibrium is achieved, we recover the eigenfrequency of the breathing mode of the stress-free nearly spherical shapes to be $f_{sf} \approx$ 2.05 MHz and 1.35 MHz for Cases I and II, respectively. Both values are very close to the analytically obtained eigenfrequencies of the breathing mode of the respective shells [34], due to the nearly spherical initial shape, as well as those reported by Chabouh et al. [39] for the same shell size and dilatational elasticity. We next repeated the above procedure to determine the resonance frequency of the compressed static shape, $f_c$, corresponding to the static equilibrium obtained when $\varepsilon$=0.8 for Case I and $\varepsilon$=0.6 for Case II, in the manner described earlier for the stress-free shape. In particular, we applied a pressure jump, $\varepsilon$

= 0.9 and $\varepsilon$ = 0.7 pertaining to Cases I and II, respectively, to the static configuration corresponding to $\varepsilon$ = 0.8 for Case I and to $\varepsilon$ = 0.6 for Case II, and recorded the period of the decaying oscillations until the new static configuration was reached. Figures 7(c), 7(d) illustrate a significant reduction of the breathing mode eigenfrequencies for Cases I and II down to $f_c \approx$ 0.37 MHz and $f_c \approx$ 0.27 MHz, respectively.

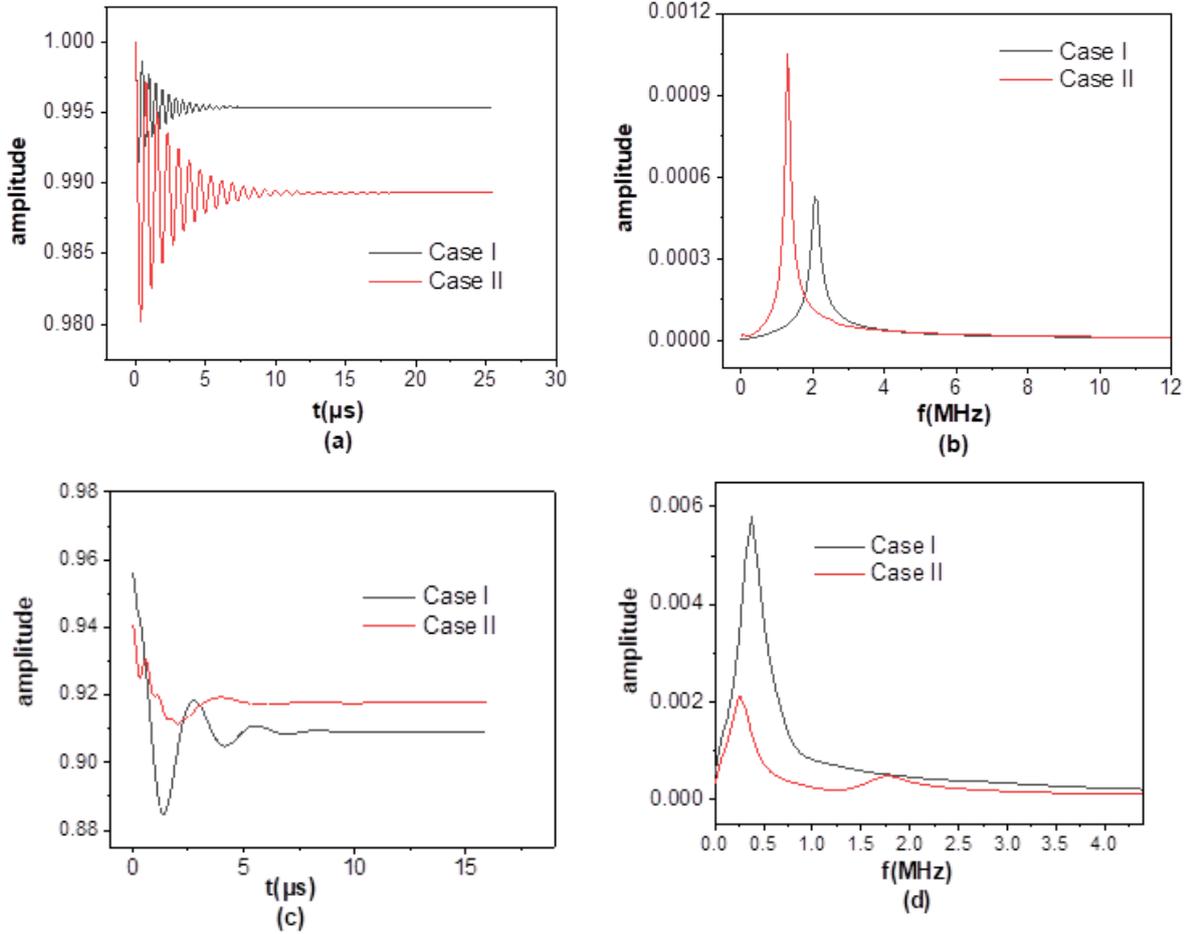

FIG. 7. (a, b) Temporal evolution and FFT analysis, respectively, of the breathing mode, $P_0$, amplitude for Cases I and II and a step change disturbance of $\varepsilon$=0.1 on the initial stress-free state. (c, d) Temporal evolution and FFT analysis, respectively, of the breathing mode, $P_0$, amplitude for Cases I and II and a step change disturbance of $\varepsilon$=0. 9 and $\varepsilon$ = 0.7 on the static configuration corresponding to $\varepsilon$ = 0.8 and $\varepsilon$ = 0.6 corresponding to Case I and Case II, respectively.

To investigate the manner in which microbubble motion, of the kind observed in experiments with fabricated microswimmers, is triggered by acoustic disturbances and to explore the potential role of shell buckling on the onset and velocity of translation, we examine the dynamic response of Case I to acoustic signals. To this end, we set the sound frequency to $f_{ext}$ = 0.5 MHz, a value significantly lower than the resonance frequency of the perfect bubble of the same size, and perform an amplitude sweep in response to an acoustic disturbance, of the form described in Eq. (1), on the stress-free shape with an imperfection on the north pole of the microbubble. In all cases presented in Figs. 8 and 9 buckling initiates at the position of the imperfection at $\theta = 0$ with the asymmetric modes $P_3$ and $P_5$ emerging alongside the symmetric mode $P_4$ from the outset as the bubble begins to undergo volume pulsations. Shortly thereafter, the bubble starts translating in the direction of concavity, as evidenced by the temporal evolution of the center of mass, $z_{cm}$, shown in Figs. 8(a)-8(c). The frequency that dominates the oscillations matches the external frequency, $f_{ext}$, as illustrated in Fig. 8(d) where the FFT analysis of the breathing mode oscillations, $P_0$, is presented. However, as the sound amplitude increases, harmonic frequencies also emerge. As expected, shape deformation intensifies with increasing sound amplitude, see Figs. 8(a)-8(c) depicting the shape mode decomposition and Fig. 9 illustrating the bubble shape evolution during maximum compression with increasing sound amplitude. The increased deformation leads to a corresponding rise in the maximum instantaneous bubble velocity, occurring immediately after maximum compression. The mean speed of translation, calculated via $u'_m \simeq \Delta z'_{cm} / \Delta t' = \omega_{ext} R_{0,eq} \left( \Delta z_{cm} / \Delta t \right)$ with $\omega_{ext} \equiv 2\pi f_{ext}$, also increases with sound amplitude in the manner presented in Fig. 8(e).

Interestingly, as shown in Fig. 9, in all cases the bubble exhibits steady shape pulsations between an almost spherical shape at maximum expansion and a buckled shape at maximum compression. The volume and shape of the bubble at maximum compression closely match the corresponding asymmetric static buckled shape predicted by the asymmetric bifurcation branch in Fig. 3(a) for the same disturbance amplitude ε. This implies that the initial imperfection dynamically drives the bubble toward the static asymmetric branch for disturbance amplitudes much smaller than the critical amplitude for static buckling ($\varepsilon_{cr} \approx 1.56$) associated with the spherical branch. This is also known as the amplitude knock down effect in the literature of shell dynamics [37]. In fact, in the absence of the initial imperfection, the bubble would undergo spherically symmetric volume oscillations for all cases shown in Figs. 8 and 9.

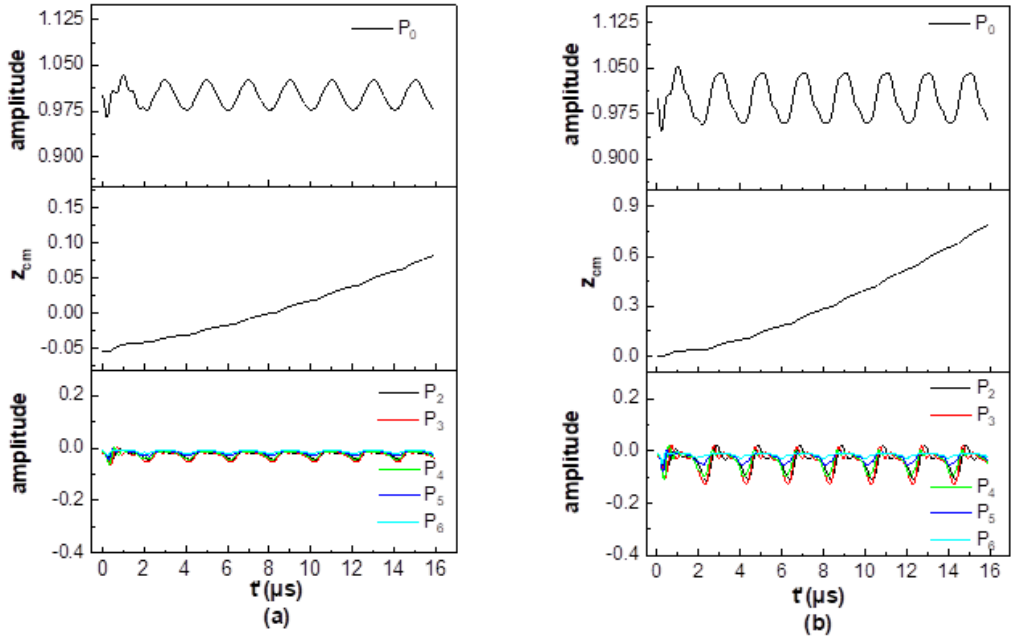

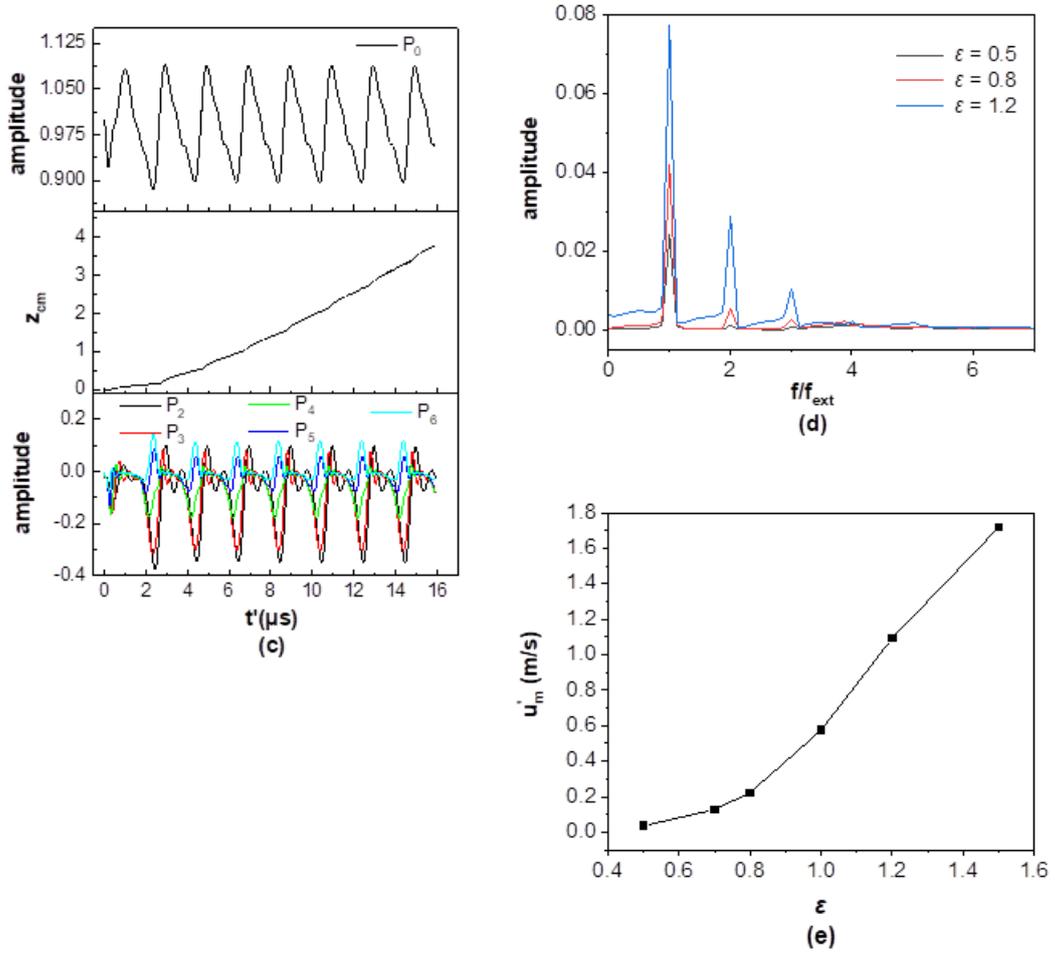

FIG. 8. (a,b,c) Temporal evolution of the breathing mode, $P_0$, center of mass, $z_{cm}$, and shape mode decomposition for sound amplitude $\varepsilon=0.5$, $\varepsilon=0.8$ and $\varepsilon=1.2$, respectively, (d) FFT analysis on the oscillations of the breathing mode, $P_0$, and (e) mean translational velocity as a function of sound amplitude, when an acoustic disturbance of frequency 0.5MHz is imposed on Case I.

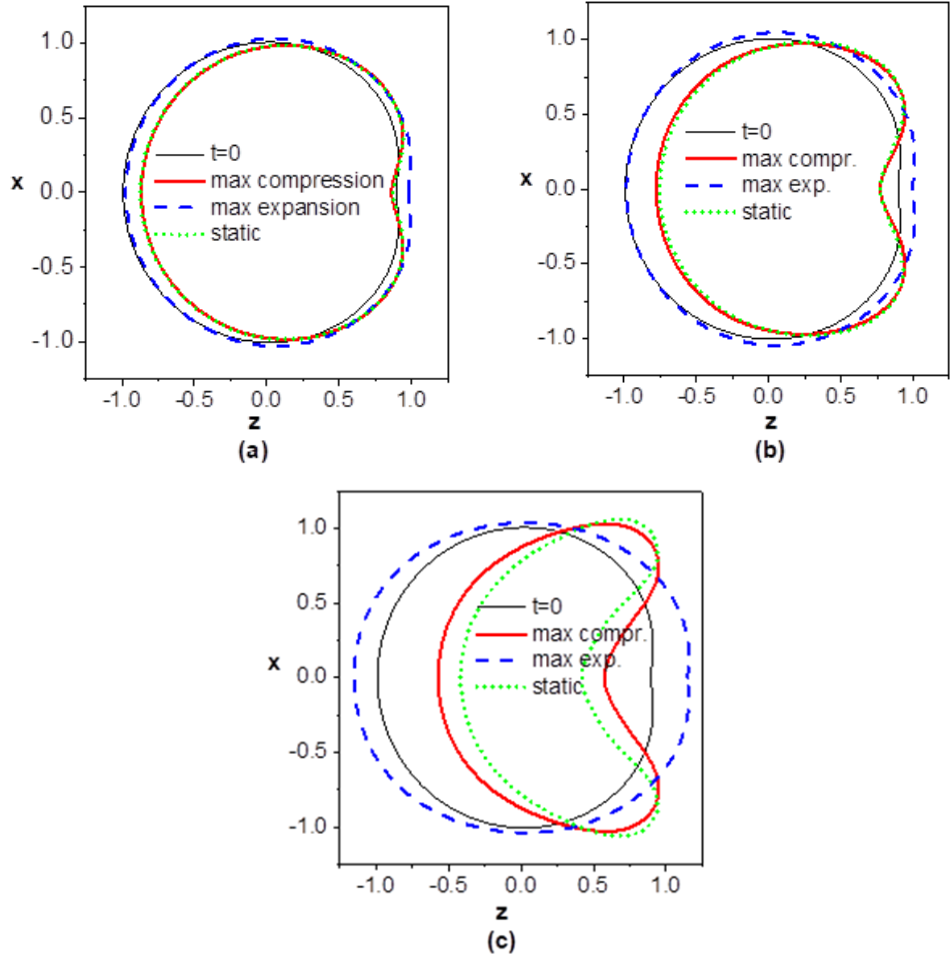

FIG. 9. Shape of the bubble at t=0, maximum compression, maximum expansion and corresponding static shape for amplitude (a) ε=0.5, (b) ε=0.8 and (c) ε=1.2, when an acoustic disturbance of frequency 0.5MHz is imposed on Case I.

Pertaining to the mechanism driving the bubble's movement, we demonstrate below that it is the same as the one identified by Vlachomitrou and Pelekasis [22] for asymmetric concave-up shapes. In that study, it was argued that when the compression-only effect results in asymmetric concave-up shapes, the translational motion of the microbubble towards the wall that lies beneath it due to the Bjerknes forces, is interrupted. This occurs due to a positive pressure drag that is generated because of the emerging buckled, asymmetric, concave upwards shapes. As a result, the microbubble begins to move in the opposite direction, remaining aligned with the direction of concavity. The same positive pressure force is observed in the present study, as illustrated in Fig. 10. In Fig. 10(a) the temporal evolution of the pressure force, $F_P$, and viscous drag, $F_V$:

$$F_p = -\iint p\mathbf{n}\cdot\mathbf{e}_z dA, \quad F_v = \frac{1}{\text{Re}}\iint \mathbf{e}_z\cdot\mathbf{n}\cdot\underline{\underline{\tau}}_l dA, \quad dA = r\sin\theta dSd\varphi, \tag{16}$$

are presented over three periods of the forcing for the simulation shown in Fig. 8(b). In Fig. 10(b), the mean values of these forces, $F_{P,av}$ and $F_{V,av}$, are calculated by averaging over each period of the forcing, confirming the same underlying translational mechanism described in [22]. At this point we should stress that, in the numerical simulations presented in Figs. 7-9, the origin of the coordinate system is shifted at regular time intervals to the midpoint of the distance between the two poles. This adjustment ensures that the poles are positioned on either side of the origin, accurately representing the bubble's shape. However, for the calculation of forces shown in Fig. 10, we kept the origin of the axes fixed over three periods of forcing, in order to clearly capture the microbubble's translation toward the concave direction as illustrated in Fig. 10(c).

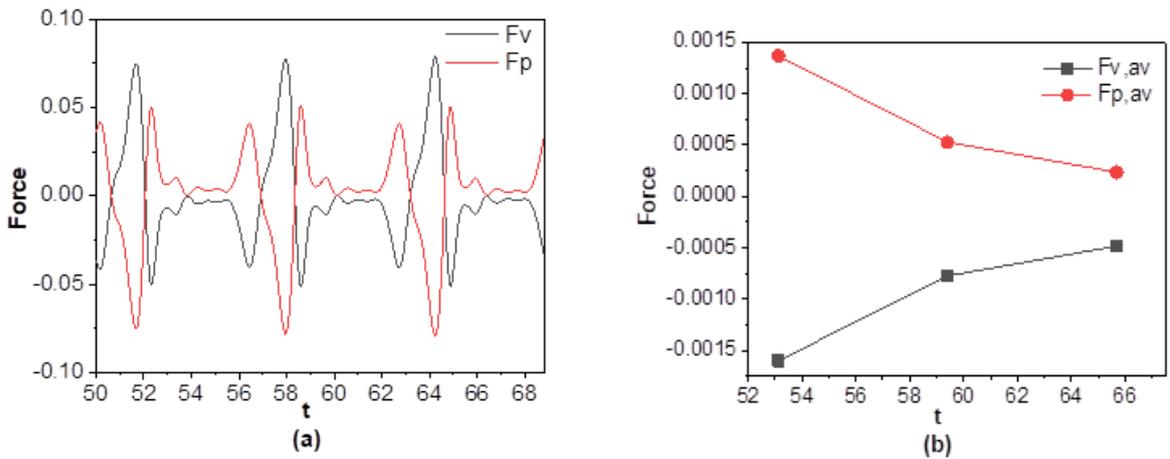

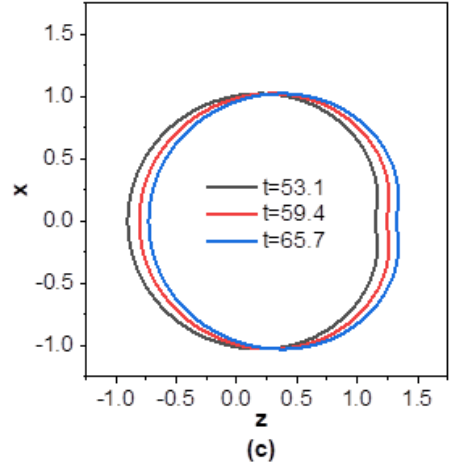

(c)

FIG. 10. Time evolution of (a) the instantaneous force on the microbubble due to pressure, $F_P$, and viscous drag, $F_V$, within 3 periods of the forcing, (b) the average force due to pressure, $F_{P,av}$, and average viscous drag, $F_{V,av}$, averaged over each period of the forcing, and (c) motion of the bubble to the right and mean shape around which it oscillates for each period, when an acoustic disturbance of frequency 0.5 MHz and amplitude ε=0.8 is imposed on Case I.

We next proceed to examine the dependence of the translational velocity on the external frequency, as well as conditions under which the speed is maximized. To this end, we set the sound amplitude to $\varepsilon = 0.8$, perform a frequency sweep from $f_{ext} = 0.1$ - 2.5 MHz and report the evolution of the speed of translation, averaged over a period of the pulsation, shown in Fig. 11. Interestingly, the optimum frequency does not coincide with the resonance frequency of the initial shape ($f_{sf} \approx 2.05$ MHz) or the resonance frequency of the compressed static shape that the bubble adopts during maximum compression when ε=0.8 ($f_c \approx 0.37$ MHz) as one might expect. The latter frequency of the compressed static shape, $f_c$, is calculated numerically in the manner described earlier; see the discussion in reference to Fig. 7. In fact, the optimal frequency is determined by the difference between these two frequencies, i.e. $f_{ext,opt} = f_{sf} - f_c = 2.05 - 0.37 \approx 1.7$ MHz. As revealed by the numerical simulations presented in Fig. 12, for $f_{ext} = f_{sf} - f_c \approx 1.7$ MHz the asymmetric mode $P_5$ becomes quite significant and eventually dominates over the lower modes $P_2$-$P_4$ which prevail for $f_{ext} = f_c$ and $f_{ext} = f_{sf}$. This results in a greater energy transfer to the translational mode, $P_1= z_{cm} - z_{cm}(t=0)$, thereby maximizing the translation speed. It should be noted that as the disturbance amplitude increases, to $O(\varepsilon^2)$ interaction and energy transfer between the two dominant time scales, i.e. those pertaining to $f_{sf}$ and $f_c$, are favored giving rise to composite frequencies that arise by adding and subtracting the above two fundamental frequencies. In this fashion, over a long time scale deformed shapes arise near maximum compression as a result of energy transfer between the nearly spherical mode corresponding to $f_{sf}$ and the asymmetric Legendre modes emerging due to $f_c$. Subsequently, interaction between consecutive shape modes triggers growth of $P_1$, thereby favoring translation in the direction of concavity in the manner previously discussed. This process is more intense when $f_{ext} = f_{sf} - f_c$ in comparison with the sum $f_{sf} + f_c$ since in the latter case viscous damping is increased due to the larger overall frequency.

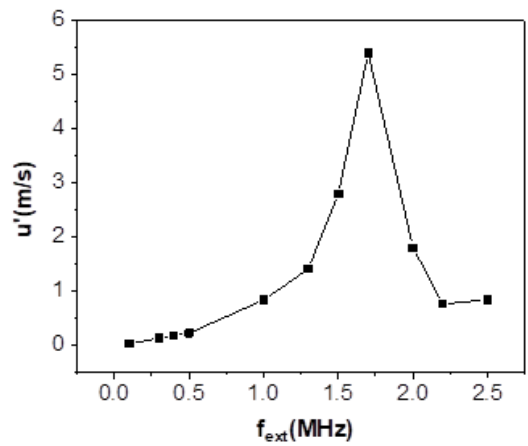

FIG. 11. Speed of bubble translation as a function of the external frequency for Case I and sound amplitude $\varepsilon = 0.8$.

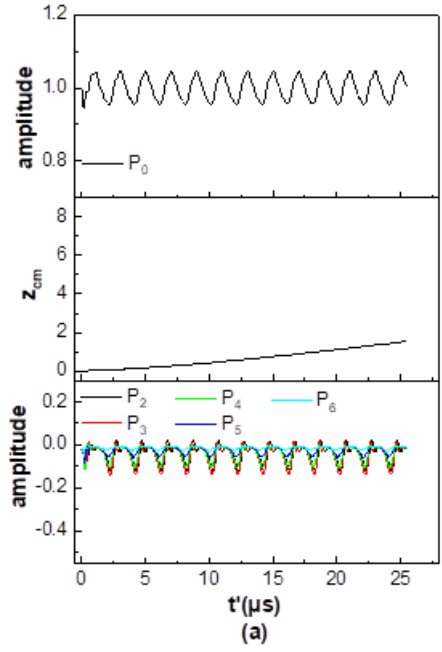
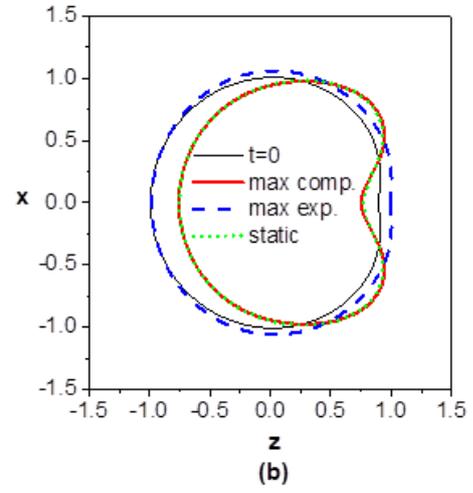

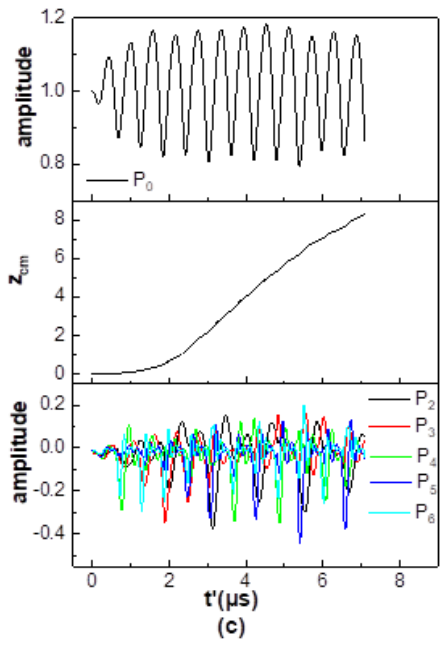
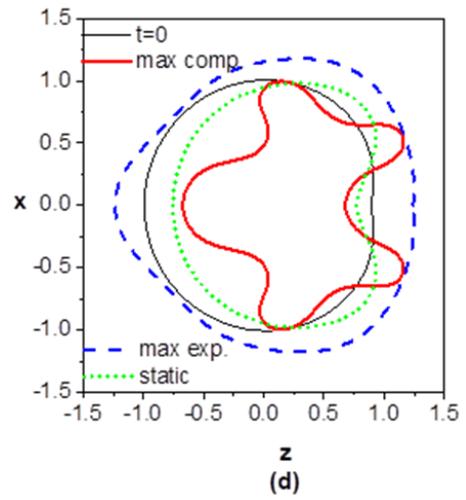

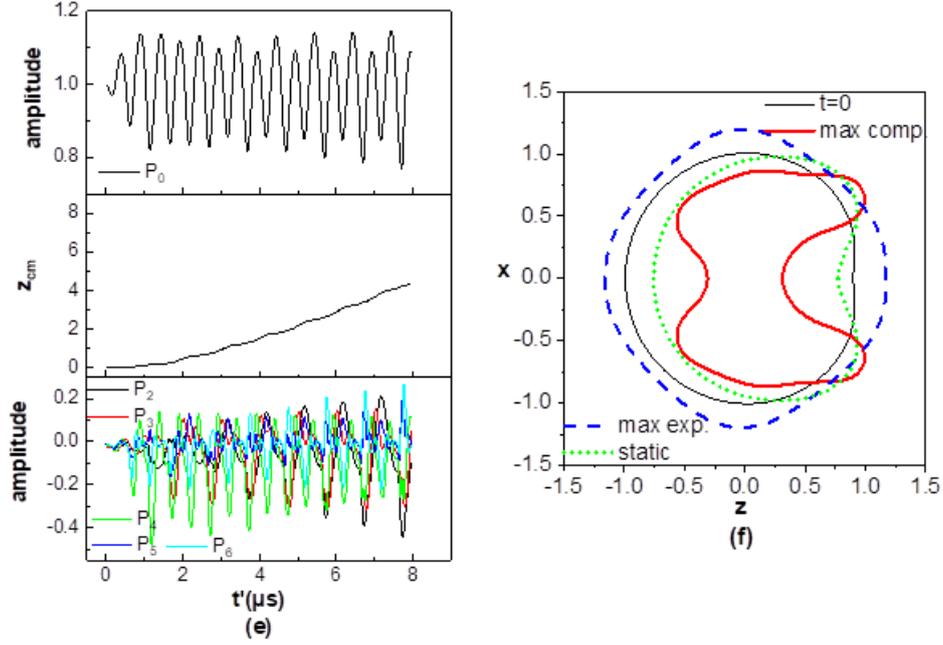

FIG. 12. (a, c, e) Temporal evolution of the breathing mode, $P_0$, center of mass, $z_{cm}$, and shape mode decomposition and (b, d, f) shape of the bubble at t=0, maximum compression, maximum expansion and corresponding static shape for Case I, sound amplitude $\varepsilon = 0.8$ and external frequency $f_{ext}$ =0.4 MHz ≈ $f_c$=0.37 MHz, 1.7 MHz≈ $f_{ext,opt}$ and 2.05 MHz≈ $f_{sf}$, respectively.

Next, we consider a contrast agent with the properties of Case II, which features a softer shell compared to Case I. To prevent bubble collapse, we set the sound amplitude to $\varepsilon = 0.6$, since Case II has a lower buckling threshold, see also Fig. 3. A frequency sweep is performed in the interval $f_{ext}$ = 0.3 - 1.5 MHz. As shown in Fig. 13, the difference between the resonance frequency of the stress-free shape ($f_{sf}$ ≈ 1.35 MHz) and the resonance frequency of the compressed static shape, corresponding to the imposed sound amplitude ($f_c$ ≈ 0.27 MHz), determines again the optimum frequency, $f_{ext,opt}$ = $f_{sf}$-$f_c$ ≈ 1.1 MHz, for maximum translational velocity. Additionally, it is evident that when the acoustic signal is tuned near the resonance frequency of the compressed static shape, $f_{ext}$=0.3 MHz ≈ $f_c$, the bubble exhibits saturated shape pulsations, slowly moving at a velocity of 0.07m/s, see Fig. 14(a). The bubble oscillates between an almost spherical shape during maximum expansion and a shape very close to the static solution of the asymmetric static branch of Fig. 3(b) during maximum compression, as illustrated in Fig. 14(b). Figure 14(a) also highlights the strain softening behavior of the shell, where the amplitude of the breathing mode, $P_0$, during expansion exceeds that during compression.

When the acoustic frequency is set close to the resonance frequency of the initial shape, $f_{ext}$=1.3 MHz ≈ $f_{sf}$, the saturated shape pulsations that the bubble undergoes become more intense, with the symmetric mode $P_4$ prevailing, as shown in Figs. 14(e), 14(f). However, asymmetric modes are also present, with the dominant mode $P_3$ being predominantly negative during one period of oscillation, leading to translation in the direction of the imperfection around the north pole.

Upon close examination of Figs. 14(c), 14(d), which correspond to the optimal frequency $f_{ext}$= 1.1 MHz ≈ $f_{sf}$ - $f_c$, shape oscillations become unsaturated and more intense resulting in highly distorted shapes, particularly during compression. In fact, when comparing against shape modes at the optimal frequency of Case I, Figs. 12(c), 12(d), Case II exhibit significantly larger shape deformations, even though the sound amplitude is smaller, due to the reduced area dilatational modulus. The interaction between these enhanced shape modes leads to a substantial transfer of energy to mode $P_1$ thereby maximizing the translation speed at $f_{ext}$ ≈ $f_{sf}$ - $f_c$, as was also observed for Case I. However, the advantage of maximizing speed must be weighed against the large interfacial deformations, which could potentially lead to bubble collapse.

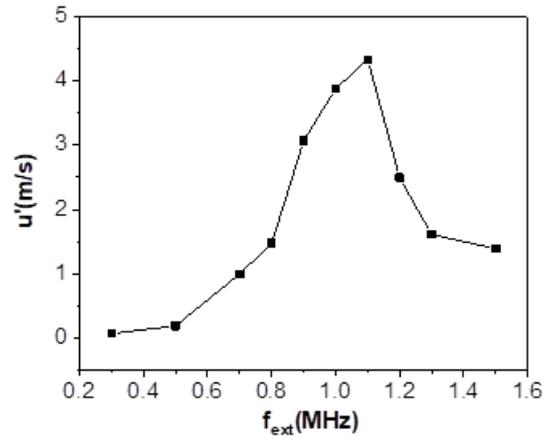

FIG. 13. Speed of bubble translation as a function of the external frequency for Case II and sound amplitude $\varepsilon = 0.6$.

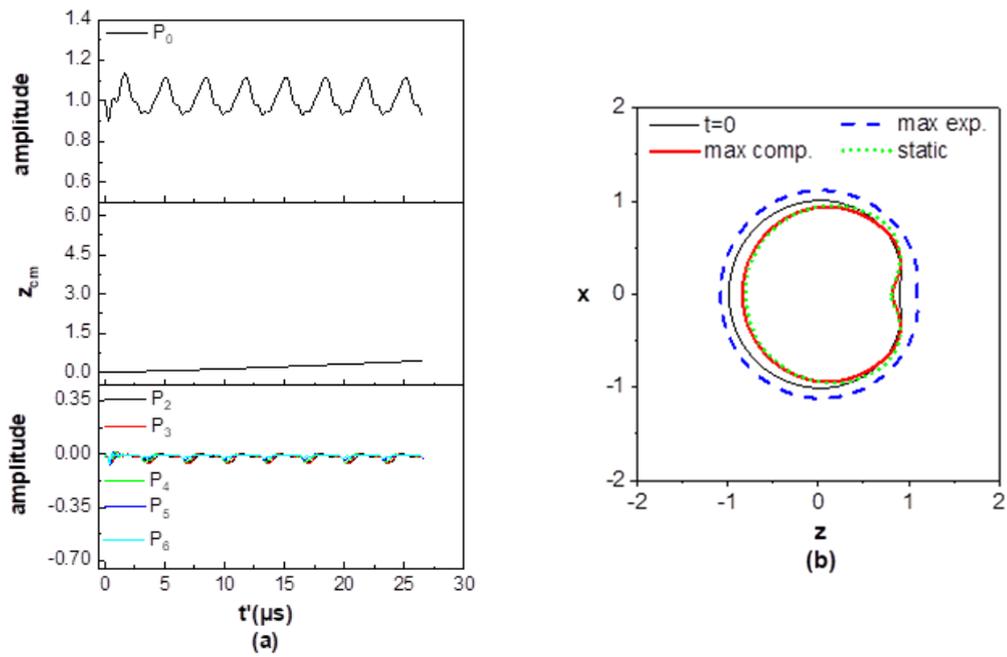

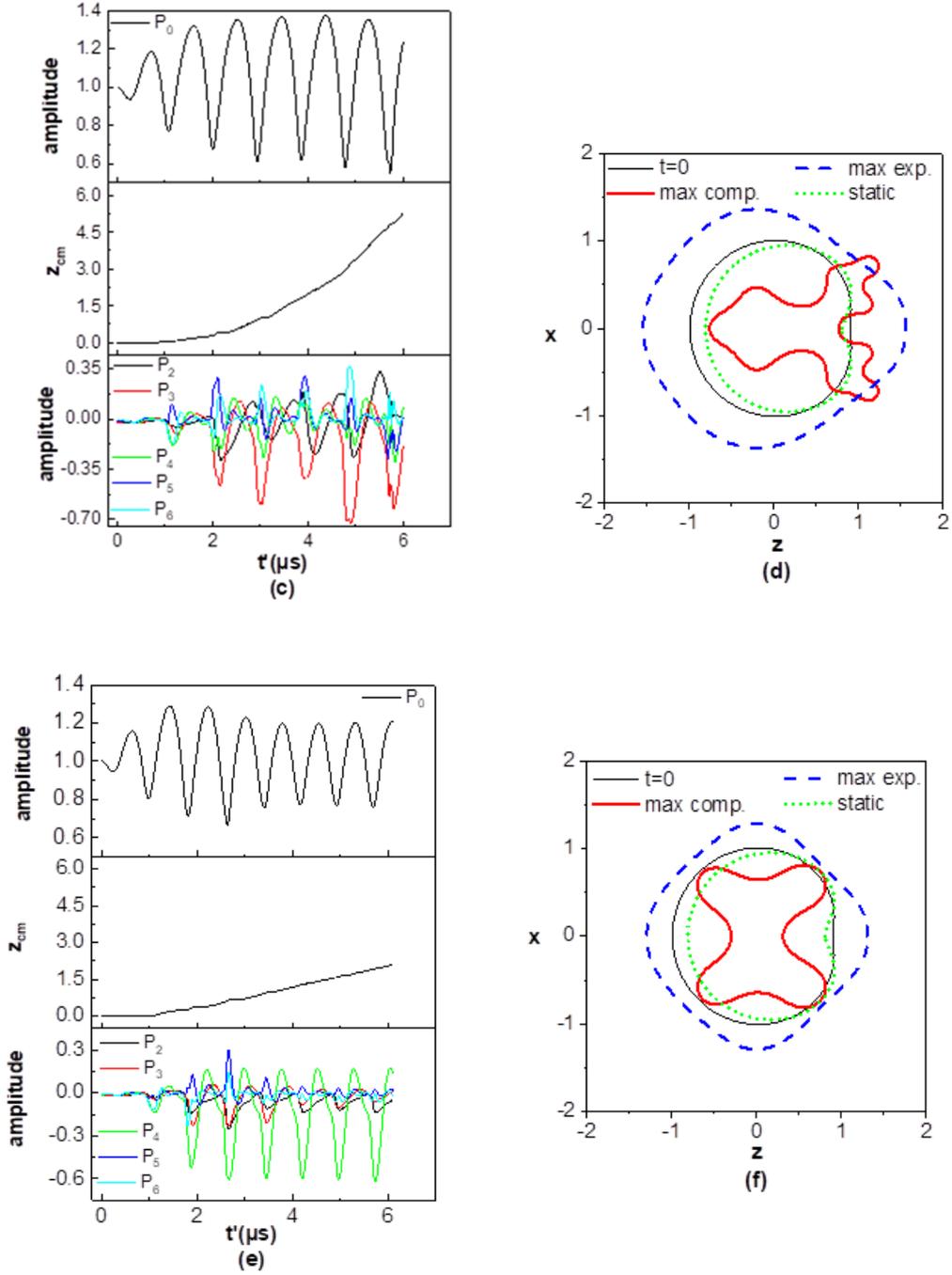

FIG. 14. (a, c, e) Temporal evolution of the breathing mode, $P_0$, center of mass, $z_{cm}$, and shape mode decomposition and (b, d, f) shape of the bubble at t=0, maximum compression, maximum expansion and corresponding static shape for Case II, sound amplitude $\varepsilon = 0.6$ and external frequency $f_{ext} = 0.3$ MHz $\approx f_c=0.27$ MHz, 1.1 MHz$\approx f_{ext,opt}$, and 1.3 MHz$\approx f_{sf}$, respectively.

## V. CONCLUSIONS

Dynamic similarities in the experiments of Chabouh et al. [11] and the simulations of Vlachomitrou & Pelekasis [22], particularly the reported asymmetric buckled shapes including assertions that such shapes dictate the direction of motion, inspired the present study. We examined encapsulated microbubbles containing a small defect in the region around one of the two poles, as reported in the above experimental study, with shell properties similar to those of the fabricated microbubbles used by Chabouh et al. [11].

Bifurcation diagrams revealed that the initial imperfection of the bubble shape significantly affects its static buckling behavior. When an indentation is introduced to the stress-free shape, the bubble experiences a spontaneous transition to the buckling branch at a lower external overpressure compared to a perfect spherical shell. Thus, the imperfection always forces the bubble to evolve along the more stable path, instead of the typical subcritical branch, without exhibiting a limit point. This transition is influenced by the nature of the initial imperfection, with asymmetric imperfections leading to an early shift to the asymmetric buckling branch, while symmetric imperfections cause a transition along the symmetric path, in a manner that is typically observed in "unfoldings" of bifurcation points. The resulting branches closely resemble those emerging from a perfect sphere after the limit points, with the static shapes being more compressed for softer shells. Dynamic simulations performed to investigate the microbubbles' behavior in an unbounded flow subject to acoustic disturbances, confirmed these findings and, moreover, revealed that the bubble adopts the corresponding static shapes during maximum compression, especially when the acoustic signal is tuned to the resonance frequency of the respective buckled shape.

Numerical simulations also captured the translational motion of the bubble. It was seen that as the bubble undergoes volume oscillations buckling occurs at the location of the initial imperfection ($\theta = 0$), while the bubble starts translating in the concave direction as shown by the temporal evolution of the center of mass, in agreement with the experiments of Chabouh et al. [11]. The motion is governed by a positive pressure force, that triggers bubble motion in the direction of the concave side of the interface. This mechanism is very similar to the one identified by Vlachomitrou and Pelekasis [22] for asymmetric concave-up shapes, where compression-only behavior is associated with shell buckling that generates a shift in the direction of the bubble translation, thus interrupting the initial translational motion towards the wall.

The translational velocity of the bubble exhibits a clear dependence on the external acoustic frequency. A frequency sweep revealed that the maximum translation speed is not obtained at the resonance frequency of the stress-free shape, $f_{sf}$, nor at the resonance frequency of the compressed static shape, $f_c$, corresponding to the imposed sound amplitude. In fact, when the frequency is tuned to the resonance frequency of the compressed static shape ($f_{ext}=f_c$) the bubble exhibits saturated shape pulsations with its shape resembling the corresponding static buckled shape during maximum compression, while moving at slower translational speeds. When the forcing frequency is set close to the resonance frequency of the stress-free shape ($f_{ext}=f_{sf}$), the bubble undergoes more intense saturated pulsations, and the speed of translation is increased. However, the optimal external frequency for maximizing translational speed is determined by the difference between the resonance frequency of the stress-free shape and the resonance frequency of the compressed static shape, $f_{ext,opt}=f_{sf}-f_c$. For Case I, this corresponds to $f_{ext,opt} \approx 1.7$MHz, while for the softer shell of Case II, the optimal frequency is $f_{ext,opt} \approx 1.1$ MHz. This optimal frequency maximizes energy transfer to the compressed shape, facilitates rapid growth of asymmetric modes along with symmetric ones, and consequently triggers the translational mode $P_1$ via nonlinear interaction of consecutive Legendre modes, thus achieving the highest observed speed that is 5.4 m/s for Case I and ε = 0.8 and 4.3 m/s for Case II and ε=0.6.

It should be stressed that the above values pertaining to the peak translational velocity and the corresponding forcing frequency are lower than those reported by Chabouh et al. [11] for microbubbles with similar size, shell elasticity and viscosity and consequently similar eigenfrequency at stress free conditions, $f_{sf}$. Based on the above proposed mechanism this is attributed to the larger eigenfrequency of the buckled shape, $f_c$, for the same sound amplitude, which amounts to a less deformed shape at maximum compression or, equivalently, to larger bending resistance. In the present study the deformation at maximum compression is larger for case I, hence the larger peak translational velocity. Nevertheless, the calculated peak velocities are quite similar for cases I and II since $k_b/(\chi R_0^2)$ is nearly the same. Owing to uncertainties in the actual bending resistance of the lipid shell involved in both cases, verification of this aspect of shell parameter impact is left to a future study.

As the sound amplitude increases or the area dilatational modulus decreases, the deformation of the bubble shape becomes more pronounced, particularly during compression. The bubble oscillates between an almost spherical shape during maximum expansion and a highly distorted shape during maximum compression. These deformations lead to higher instantaneous velocities right after maximum compression. However, as the amplitude increases or the area dilatational modulus decreases, the bubble also experiences higher strain softening behavior, where the amplitude of the breathing mode $P_0$ is larger during expansion than during compression. This non-linear deformation behavior, combined with the enhanced energy transfer to the translational mode $P_1$, results in an increase in the mean translation speed, though it also increases the risk of bubble collapse, particularly when the deformation exceeds the relevant thresholds. The entire process is a combination of pulling and pushing the surrounding fluid, after the microbubble assumes the nearly spherical and buckled states, respectively, in order to achieve translation in the direction of concavity.

While the bubble's translational speed increases with sound amplitude and the optimal external frequency, the large interfacial deformations that accompany these motions pose a significant risk of bubble collapse. This highlights the trade-off between maximizing translational velocity and maintaining the bubble's integrity. Careful consideration of the shell parameters is necessary during fabrication of microswimmer particles so that, combined with the appropriate acoustic excitation parameters depending on the desired application, the translational motion of the microbubble is optimized while avoiding collapse.